\documentclass[10pt]{article}
\usepackage{amsfonts, amssymb, amscd}
\usepackage{amsfonts, amssymb, amscd}
\usepackage{amsmath}
\usepackage{amssymb}
\usepackage{amsthm}
\usepackage{makeidx}
\usepackage{latexsym}
\usepackage{graphicx}

\begin{document}

\title{Modified Korteweg-de Vries surfaces}

\vspace{5cm}

\author{ S{\" u}leyman Tek\\
{\small  Department of Mathematics, Faculty of Science}\\
{\small Bilkent University, 06800 Ankara, Turkey}\\}

\vspace{5cm}

\begin{titlepage}
\maketitle

\begin{abstract}
In this work, we consider $2$-surfaces in ${\mathbb R}^3$ arising
from the modified Korteweg de Vries  (mKdV) equation. We give a
method for constructing the position vector of the mKdV surface
explicitly for a given solution of the mKdV equation. mKdV
surfaces contain Willmore-like and Weingarten surfaces. We show
that some mKdV surfaces can be obtained from a variational
principle where the Lagrange function is a polynomial of the
Gaussian and mean curvatures.
\end{abstract}

\end{titlepage}

\section{Introduction}
 In this work we study the 2-surfaces in
${\mathbb R}^3$ arising from the deformations of the modified
Korteweg de Vries (mKdV) equation and its Lax pair. Deformation
technique was developed by several authors. Here we mainly follow
references \cite{sym1}-\cite{do}.

 Let $u(x,t)$ satisfy the mKdV equation
 \begin{equation}{\label{mKdV0}}
u_{t}=u_{3x}+\frac{3}{2}u^2u_{x}.
 \end{equation}

\noindent Substituting the traveling wave ansatz $u_{t}-\alpha\,
u_{x}=0$ in Eq. (\ref{mKdV0}), where $\alpha$ is an arbitrary real
constant, we get
\begin{equation}{\label{mKdVequation}}
  u_{2x}=\alpha u-\frac{u^3}{2}.
\end{equation}
\noindent Here and in what follows, subscripts $x$, $t,$ and
$\lambda$ denote the derivatives of the objects with respect to
$x$, $t,$ and $\lambda$, respectively. The subscript $nx$ stands
for $n$ times $x$ derivative, where $n$ is a positive integer,
e.g., $u_{2x}$ indicates the second derivative of $u$ with respect
to $x$. We use Einstein's summation convention on repeated indices
over their range. Eq. (\ref{mKdVequation}) can be obtained from a
Lax pair $U$ and $V$, where
 \begin{eqnarray}{\label{U mKdV}}
 U&=&{\frac{i}{2}}\left(%
\begin{array}{cc}
  \lambda & -u \\
  -u & -\lambda \\
\end{array}%
\right),
\end{eqnarray}
 \begin{eqnarray}{\label{V mKdV}}
 V&=&-{\frac{i}{2}}
     \left(%
          \begin{array}{cc}
         \frac{1}{2}u^2-(\alpha+\alpha \lambda+\lambda^2) &
          (\alpha+\lambda)u-iu_{x} \\
          (\alpha+\lambda)u+iu_{x} &
        -\frac{1}{2}u^2+(\alpha+\alpha \lambda+\lambda^2) \\
        \end{array}%
    \right),
\end{eqnarray}
and $\lambda$  is the spectral parameter. The Lax equations are
given as
\begin{equation}
 \Phi_{x}=U\, \Phi~~~,~~~\Phi_{t}=V\, \Phi , \label{cur2}
\end{equation}
where the integrability of these equations are guaranteed by the
mKdV equation or the zero curvature condition
\begin{equation}
U_{t}-V_{x}+[U,V]=0. \label{cur1}
\end{equation}

\noindent A connection of the mKdV equation to surfaces in
${\mathbb R}^3$ can be achieved by defining $su(2)$ valued $2
\times 2$ matrices $A$ and $B$ satisfying

\noindent
\begin{equation}
A_{t}-B_{x}+[A,V]+[U,B]=0 \label{eq1}.
\end{equation}

\noindent Let $F$ be an $su(2)$ valued position vector of the
surface $S$ corresponding to the mKdV equation such that

\begin{equation}
y_{j}=F_{j}(x,t;\lambda)~~ ,~j=1,2,3~~ ,~~F= i\,\sum^{3}_{k=1}
F_{k}\, \sigma_{k}, \label{eq2}
\end{equation}
where $\sigma_{k}$'s are the Pauli sigma matrices

\begin{equation}
\sigma_{1}=\left( {\begin{array}{ll}
                           0 & 1 \\
                            1& 0
                            \end{array}}
                            \right),~~
  \sigma_{2}=\left( {\begin{array}{ll}
                           0 & -i \\
                            i & 0
                            \end{array}}
                            \right),~~
\sigma_{3}=\left( {\begin{array}{ll}
                           1 & 0 \\
                            0 & -1
                            \end{array}}
                            \right).
\end{equation}
The connection formula (connecting integrable systems to
2-surfaces in ${\mathbb R}^3$) is given by

 \begin{equation}
F_{x}=\Phi^{-1}\,A\, \Phi~~,~~ F_{t}=\Phi^{-1}\,B\, \Phi.
\label{eq0}
 \end{equation}

\noindent Then at each point on $S$, there exists a frame $\{
F_{x}, F_{t}, \Phi^{-1} C \Phi \}$ forming a basis of ${\mathbb
R}^3$, where $C={[A,B]}/{||[A,B]||}$ and $[A,B]$ denotes the usual
commutator $[A,B]=AB-BA$. The inner product $<,>$ of $su(2)$
valued vectors $X$ and $Y$ are given by $<X,Y>=-{1 \over 2} \,\,
\mbox{trace} (XY)$. Hence $||X||=\sqrt{|<X,X>|}$. The first and
second fundamental forms of $S$ are
\begin{eqnarray}
&&(ds_{I})^2 \equiv g_{ij}\,dx^{i}\,dx^{j}=
<A,A>\, dx^2+2<A,B>\,dx\,dt+<B,B>\,dt^2 , \nonumber\\
&&(ds_{II})^2 \equiv h_{ij}\,dx^{i}\,dx^{j} =<A_{x}+[A,U],C>\,
dx^2 \nonumber\\
&&~~~~~~~~~~~~+2<A_{t}+[A,V],C>\,dx\,dt +<B_{t}+[B,V],C>\,dt^2,
\label{eq5}
\end{eqnarray}
\noindent where $i,j=1,2$, $x^1=x$ and $x^2=t$. Here $g_{ij}$ and
$h_{ij}$ are coefficients of the first and second fundamental
forms, respectively. The Gauss and the mean curvatures of $S$ are,
respectively, given by $K=\mbox{det}(g^{-1}\,h)$ and
$H=\frac{1}{2}\,\, \mbox{trace}(g^{-1}\,h)$, where $g$ and $h$
denote the matrices $(g_{ij})$ and $(h_{ij})$, and $g^{-1}$ stands
for the inverse of the matrix $g$.

\vspace{0.3cm}

In order to calculate the fundamental forms in Eq. (\ref{eq5}) and
the curvatures $K$ and $H$, one needs the deformations matrices
$A$ and $B$. Given $U$ and $V$, finding $A$ and $B$ from Eq.
(\ref{eq1}) is a difficult task in general. However, there are
some deformations which provide $A$ and $B$ directly. They are
given as follows:
\begin{itemize}
    \item Spectral parameter $\lambda$ invariance of the equation:
     \begin{equation}
     A=\mu\,\frac{\partial U}{\partial \lambda},~B=\mu\,\frac{\partial V}{\partial \lambda},~
     F=\mu\,\Phi^{-1}\frac{\partial \Phi}{\partial \lambda},
     \end{equation}
     where $\mu$ is an arbitrary function of $\lambda$.
    \item Symmetries of the (integrable) differential equations:
     \begin{equation}
     A=\delta U,~B=\delta V,~F=\Phi^{-1}\delta \Phi,
     \end{equation}
     where $\delta$ represents the classical Lie symmetries and (if
     integrable) the generalized symmetries of the nonlinear PDE's.
    \item Gauge symmetries of the Lax equation:
     \begin{equation}
     A=M_{x}+[M,U],~ B=M_{t}+[M,V],~F=\Phi^{-1}M\Phi,
     \end{equation}
     where $M$ is any traceless $2\times2$ matrix.
\end{itemize}

\vspace{0.3cm}

There are some surfaces which may be obtained from a variational
principle. For this purpose, we consider a functional ${\cal F}$
which is defined by

\begin{equation}
{\cal F}\equiv\int_{S}\, {\cal E}(H,K) dA+p \int_{V} dV,
\end{equation}

\noindent where ${\cal E}$ is some function of the curvatures $H$
and $K$, $p$ is a constant  and $V$ is the volume enclosed by the
surface $S$. For open surfaces, we let $p=0$. The first variation
of the functional ${\cal F}$  gives the following Euler-Lagrange
equation for the Lagrange function ${\cal E}$
\cite{tu1}-\cite{tu3}
\begin{equation}\label{el1}
(\nabla^2 +4H^2-2K) {\partial {\cal E} \over \partial
H}+2({\nabla} \cdot \bar{\nabla}+2 K H) {\partial {\cal E} \over
\partial K}-4H {\cal E} +2p=0,
\end{equation}
 where $\nabla^2$ and $\nabla{\cdot}{\bar{\nabla}}$ are defined as
\begin{equation}
 \nabla^2=
 {\frac{1}{\sqrt{\tilde{g}}}}{\frac{\partial}{\partial x^i}}
 \left({\sqrt{\tilde{g}}}{g^{ij}}{\frac{\partial}{\partial
 x^j}}\right),~\nabla{\cdot}{\bar{\nabla}}=
 {\frac{1}{\sqrt{\tilde{g}}}}{\frac{\partial}{\partial x^i}}
 \left({\sqrt{\tilde{g}}}K{h^{ij}}{\frac{\partial}{\partial
 x^j}}\right),
 \end{equation}
 and ${\tilde{g}}={\det{(g_{ij})}}$, where $g^{ij}$ and $h^{ij}$ are
 the inverse components of the first and second fundamental forms,
 $x^{1}=x, x^2=t$. The following are examples of surfaces derived from
 a variational principle:

\begin{itemize}
    \item [{\bf i)}]Minimal surfaces: $ {\cal E}=1, ~~p=0$;
    \item [{\bf ii)}]constant mean curvature surfaces: ${\cal
    E}=1$;
    \item [{\bf iii)}]linear Weingarten surfaces: ${\cal E}=aH+b$, where $a$
and $b$ are some constants;
    \item [{\bf iv)}]Willmore surfaces: $ {\cal E}=H^2$ \cite{will1},
    \cite{will2};
    \item [{\bf v)}]surfaces solving the shape equation of lipid membrane: ${\cal E}=(H-c)^2$,
where $c$ is a constant \cite{tu1}-\cite{tu3},
\cite{hel1}-\cite{mlad};
\end{itemize}

\vspace{0.3cm}

\noindent The surfaces obtained from the solutions of the equation
\begin{equation}{\label{Willmore-like}}
      \nabla^2H+{a}H^3+bH\,K=0,
\end{equation}
are called {\it Willmore-like} surfaces, where $\nabla^2$ is the
Laplace-Beltrami operator defined on the surface and $a$, $b$ are
arbitrary constants.

\vspace{0.3cm} \noindent
 Unless $a=2$ and $b=-2$, these
surfaces do not arise from a variational problem. The case
$a=-b=2$ corresponds to the Willmore surfaces.
 For compact 2-surfaces, the constant $p$ may be different than
zero, but for noncompact surfaces we assume it to be zero. For the
latter, we require asymptotic conditions, where $K$ goes to a
constant and $H$ goes to zero. This requires that the  mKdV
equation have solutions decaying rapidly to zero as $|x|
\rightarrow \pm \infty$. Soliton solutions of the mKdV equation
satisfy this requirement. In this work, using solitonic solutions
of the mKdV equation, we find the corresponding $2$-surfaces and
then solve the Euler-Lagrange equation [Eq. (\ref{el1})] for
polynomial Lagrange functions of $H$ and $K$, i.e.
\begin{equation}
{\cal
E}=a_{N}\,H^{N}+\ldots+b_{10}\,K\,H+b_{11}\,K\,H^{2}+\ldots+e_{1}\,K+...
 \end{equation}
 For each $N$, we find the constants $a_{l},$ $b_{nk},$ and $e_{m}$ in
terms of others and the parameters of the surface.

From a solution of the mKdV equation, we first find the
fundamental forms in Eq. (\ref{eq5}) and the curvatures $K$ and
$H$ of the corresponding $2$-surface $S$. To find the position
vector $\overrightarrow{y}(x,t)$ of $S$, we use Eq. (\ref{eq0}).
To solve this equation, we need the matrix $\Phi$ satisfying the
Lax equation [Eq. (\ref{cur2})] for a given function $u(x,t)$.
Hence, in general our method for constructing the position vector
$\overrightarrow{y}$ of integrable surfaces consists of the
following steps:

 \vspace{0.3 cm}

 {\noindent}{\bf{i)}} Find a solution $u(x,t)$ of the mKdV
 equation.

\vspace{0.3 cm}

{\noindent}{\bf{ii)}} Find a solution of the Lax equation [Eq.
(\ref{cur2})] for a given $u(x,t)$.

\vspace{0.3 cm}

{\noindent}{\bf{iii)}} Find the corresponding deformation matrices
$A$, $B$, and find $F$ from Eq. (\ref{eq0}).

In this work more specifically, starting  with \textit{one
soliton} solution of the mKdV equation and following the steps
above, we solve the Lax equations and find the corresponding
$SU(2)$ valued function $\Phi(x,t)$. Then using the spectral
deformations and combination of the gauge and spectral
deformations, we find the parametric representations (position
vectors) of the mKdV surfaces and plot some of them for some
special values of constants. We show that there are some
Weingarten and Willmore-like mKdV surfaces obtained from spectral
deformations. Surfaces arising from a combination of the gauge and
spectral deformations do not contain Willmore-like surfaces. We
study also the mKdV surfaces corresponding to the symmetry
deformations. We determine all geometric quantities in terms of
the function $u(x,t)$ and the symmetry $\phi(x,t)$. For the
simplest symmetry $\phi=u_{x}$, the surface turns out to be the
surface of the sphere with radius $|(\alpha\,\mu)/(2\,\lambda)|$,
where $\lambda$ is the spectral parameter and $\alpha$ and $\mu$
are constants.

\section{mKdV Surfaces From Spectral Deformations}{\label{mKdV
Surface}}

 In this section, we find surfaces arising from the spectral deformation
 of Lax pair for the mKdV equation. We start with the following
 proposition.

\vspace{0.3cm}

\noindent {\it{{Proposition 1:}}{\label{mKdV}}
 \,Let $u$ satisfy (which describes a traveling mKdV
 wave) Eq. (\ref{mKdVequation}). The corresponding ${{su}}(2)$
 valued Lax pair $U$ and $V$ of the mKdV equation are given by Eqs. (\ref{U
mKdV}) and (\ref{V mKdV}), respectively. Then, ${{su}}(2)$ valued
matrices $A$ and $B$ are
   \begin{eqnarray}{\label{A mKdV}}
         A&=&{\frac{i}{2}}\left(%
              \begin{array}{cc}
                  \mu & 0 \\
                   0 & -\mu \\
               \end{array}%
                \right),\\
            {\label{B mKdV}}
          B&=&-\frac{i}{2}\left(%
                   \begin{array}{cc}
                  -(\alpha\,\mu+2\,\mu\,\lambda) & \mu\,u \\
                  \mu\,u & \alpha\,\mu+2\,\mu\,\lambda \\
                    \end{array}%
                         \right),
   \end{eqnarray}
  where $A=\mu\,U_{\lambda},$
  $B=\mu\,V_{\lambda}$, $\mu$ is a constant and $\lambda$ is the
 spectral parameter. The surface $S$, generated by $U,V,A$ and $B$,
 has the following first and second fundamental forms ($j,k=1,2$):
 \begin{eqnarray}
       (ds_{I})^2&=&g_{jk}\,dx^{j}\,dx^{k}=
       \frac{\mu^2}{4}\big([dx+(\alpha+2\,\lambda)dt]^2+
       u^2\,dt^2\big),\\
        (ds_{II})^2&=&h_{jk}\,dx^{j}\,dx^{k}=
        \frac{\mu\,u}{2}\big(dx+(\alpha+\lambda)dt\big)^2+
        \frac{\mu\,u}{4}(u^2-2\,\alpha)dt^2,
 \end{eqnarray}
 with the corresponding Gaussian and mean curvatures
 \begin{equation}
 K=\frac{2}{\mu^2}\big(u^2-2\,\alpha \big),~~~
 H=\frac{1}{2\mu \,u}\big(3\,u^2+
 2\,(\lambda^2-\alpha)\big),
 \end{equation}
 where $x^1=x,$ $x^2=t.$
 }

\vspace{0.3 cm}

\noindent By using $U, V, A,$ and $B$ and the method given in the
Introduction, Proposition 1 provides the first and second
fundamental forms, and the Gaussian and mean curvatures of the
surface corresponding to spectral deformation. The following
proposition gives a class of surfaces which are Willmore-like.

\vspace{0.3 cm}

\noindent {\it{{Proposition 2:}}\, Let
$u_{x}^2=\alpha\,u^2-u^4/4$. Then the surface $S$, defined in
Proposition 1, is a Willmore-like surface, i.e., the Gaussian and
 mean curvatures satisfy Eq. (\ref{Willmore-like}),
  where
  \begin{eqnarray}
      a=\frac{4}{9},~~ b=1,~~ \alpha=\lambda^2,
  \end{eqnarray}
and $\lambda$ is an arbitrary constant.}

\vspace{0.3cm}

 It is important to search for mKdV surfaces arising
from a variational principle \cite{tu1}-\cite{tu3}. For this
purpose, we do not need a parametrization of the surface. The
fundamental forms and the Gauss and mean curvatures are enough to
look for such mKdV surfaces. The following proposition gives a
class of mKdV surfaces that solves the Euler-Lagrange equation
[Eq. (\ref{el1})].

\vspace{0.3cm}

\noindent {\it{{Proposition 3:}}\, Let
$u_{x}^2=\alpha\,u^2-u^4/4$. Then there are mKdV surfaces defined
in Proposition 1 satisfying the generalized shape equation [Eq.
(\ref{el1})] when $\cal E$ is a polynomial function of $H$ and
$K$.}

\vspace{0.3cm}

Here are several examples:

  \noindent{\it{Example 1:}} Let deg$({\cal
E})=N,$ then
\begin{itemize}
    \item [\textrm{i})]for $N=3:$\\
    ${\cal E}=a_{1}\,H^3+a_{2}\,H^2+a_{3}\,H+a_{4}+a{_5}\,K+a_{6}\,K\,H,$
    \subitem$\alpha=\lambda^2$,~$a_{1}=-\displaystyle\frac{p\,\mu^4}{72\,\lambda^4}$,~
    $a_{2}=a_{3}=a_{4}=0,$~$a_{6}=\displaystyle\frac{p\,\mu^4}{32\,\lambda^4}$,\\
    where $\lambda\neq0$, and $\mu,$ $p$, and $a_{5}$
    are arbitrary constants;

    \vspace{0.3cm}

    \item [\textrm{ii})]for $N=4:$\\
     ${\cal E}=a_{1}\,H^4+a_{2}\,H^3+a_{3}\,H^2+a_{4}\,H+a_{5}+a_{6}\,K
     +a_{7}\,K\,H+a_{8}\,K^2+a_{9}\,K\,H^2,$
    \subitem$\alpha=\lambda^2$,~$a_{2}=-\displaystyle\frac{p\,\mu^4}{72\,\lambda^4}$,~
    $a_{3}=-\displaystyle\frac{8\,\lambda^2}{15\,\mu^2}\left(27\,a_{1}-8\,a_{8}\right)$,~
    $a_{4}=0$,
    \vspace{0.1 cm}
    \subitem
    $a_{5}=\displaystyle\frac{\lambda^4}{5\,\mu^4}\left(81\,a_{1}+16\,a_{8}\right)$,~
    $a_{7}=\displaystyle\frac{p\,\mu^4}{32\,\lambda^4}$,~
    $a_{9}=-\displaystyle\frac{1}{120}\left(189\,a_{1}+64\,a_{8}\right)$,\\
    where $\lambda\neq0$, $\mu\neq0,$ and $p, a_{1}, a_{6}$, and $a_{8}$
    are arbitrary constants;

    \vspace{0.3cm}

    \item [\textrm{iii})] for $N=5:$\\
    ${\cal E}=a_{1}\,H^5+a_{2}\,H^4+a_{3}\,H^3+a_{4}\,H^2+a_{5}\,H+a_{6}+a_{7}\,K+a_{8}\,K\,H
    +a_{9}\,K^2\\
    ~~~~~~+a_{10}\,K\,H^2+a_{11}\,K^2\,H+a_{12}\,K\,H^3$,
    \subitem $\alpha=\lambda^2,$~
    $a_{3}=-\displaystyle\frac{1}{504\,\mu^2\,\lambda^4}
    \left(\lambda^6[4212\,a_{1}+256\,a_{11}]+7\,p\,\mu^6\right)$,
    \vspace{0.1 cm}
    \subitem $a_{4}=-\displaystyle
    \frac{8\,\lambda^2}{15\,\mu^2}\left(27\,a_{2}-8\,a_{9}\right),$~
    $a_{5}=\displaystyle
    \frac{6\,\lambda^4}{7\,\mu^4}\left(135\,a_{1}-88\,a_{11}\right)$,
    \vspace{0.1 cm}
     \subitem $a_{6}=\displaystyle
     \frac{\lambda^4}{5\,\mu^4}\left(81\,a_{2}+16\,a_{9}\right),$~
     $a_{8}=\displaystyle\frac{1}{32\,\mu^2\,\lambda^4}
    \left(\lambda^6[-324\,a_{1}+512\,a_{11}]+p\,\mu^6\right)$,
    \subitem $a_{10}=-\displaystyle
    \frac{1}{120}\left(189\,a_{2}+64\,a_{9}\right),$~
    $a_{12}=-\displaystyle
    \frac{1}{756}\left(1053\,a_{1}+512\,a_{11}\right)$,\\
    where $\lambda\neq0,$ $\mu\neq0,$ and $p, a_{1},$ $a_{2}, a_{7}, a_{9},$ and $a_{11}$
     are arbitrary constants;

     \vspace{0.3cm}

     \item [\textrm{iv})] for $N=6:$\\
    ${\cal E}=a_{1}\,H^6+a_{2}\,H^5+a_{3}\,H^4+a_{4}\,H^3+a_{5}\,H^2+a_{6}\,H+a_{7}+a_{8}\,K
    +a_{9}\,K\,H\\
    ~~~~~~+a_{10}\,K^2+a_{11}\,K\,H^2+a_{12}\,K^2\,H+a_{13}\,K\,H^3+a_{14}\,K^3
    +a_{15}\,K^2\,H^2\\
    ~~~~~~+a_{16}\,K\,H^4$,
    \subitem $\alpha=\lambda^2,$
    \vspace{0.1 cm}
    \subitem $a_{4}=-\displaystyle\frac{1}{504\,\mu^2\,\lambda^4}
    \left(\lambda^6[4212\,a_{2}+256\,a_{12}]+7\,p\,\mu^6\right)$,
     \vspace{0.1 cm}
    \subitem $a_{5}=-\displaystyle\frac{\lambda^4}{900\,\mu^4}
    \left(-359397\,a_{1}+191488\,a_{14}-203472\,a_{16}\right)\\
    ~~~~~~~~~~~~-\frac{8\lambda^2}{15\mu^2}
    \left(27a_{3}-8a_{10}\right)$,
    \vspace{0.1 cm}
    \subitem $a_{6}=\displaystyle\frac{6\,\lambda^4}{7\,\mu^4}
    \left(135\,a_{2}-88\,a_{12}\right),$
    \vspace{0.1 cm}
    \subitem $a_{7}=\displaystyle\frac{\lambda^6}{25\,\mu^6}
    \left(29889\,a_{1}-9856\,a_{14}+11664\,a_{16}\right)\\
    ~~~~~~~~~~~~+\frac{\lambda^4}{5\mu^4}\left(81a_{3}+16a_{10}\right)$,
    \vspace{0.1 cm}
    \subitem $a_{9}=\displaystyle\frac{1}{32\,\mu^2\,\lambda^4}
    \left(\lambda^6[-324\,a_{2}+512\,a_{12}]+p\,\mu^6\right)$,
     \vspace{0.1 cm}
     \subitem $a_{11}=-\displaystyle\frac{\lambda^2}{1800\,\mu^2}
    \left(59778\,a_{1}-13312\,a_{14}+23328\,a_{16}\right)\\
    ~~~~~~~~~~~~-\frac{1}{120}\left(189a_{3}+64a_{10}\right)$,
    \vspace{0.1 cm}
    \subitem $a_{13}=-\displaystyle\frac{1}{756}
    \left(1053\,a_{2}+512\,a_{12}\right),$
    \vspace{0.1 cm}
    \subitem $a_{15}=-\displaystyle\frac{1}{2880}
    \left(5103\,a_{1}+2048\,a_{14}+3888\,a_{16}\right),$\\
    \vspace{0.1 cm}
    where $\lambda\neq0,$ $\mu\neq0,$ and $p, a_{1},$ $a_{2}$, $a_{3}, a_{8}, a_{10},
    a_{12}, a_{14}$, and $a_{16}$
     are arbitrary constants;
 \end{itemize}
     \vspace{0.3 cm}

For general $N\geq 3$, from the above examples, the polynomial
   function $\cal E$ takes the form

   $${\cal E}=\sum_{n=0}^{N}H^{n}\sum_{l=0}^{\lfloor\frac{(N-n)}{2}\rfloor} a_{nl}K^{l},$$
   where $\lfloor x \rfloor$ denotes the greatest integer less
   than or equal to $x$ and $a_{nl}$ are constants.

\vspace{0.3 cm}
\subsection{The Parametrized form of the three parameter family of mKdV Surfaces}

 \vspace{0.3 cm}

 In the previous section, we found possible mKdV surfaces satisfying
 certain equations. In this section, we find the position vector
\begin{equation}
 \overrightarrow{y}=\left(y_{1}(x,t),y_{2}(x,t),y_{3}(x,t)\right)
 \end{equation}
 of the mKdV surfaces for a given solution of the mKdV equation and the
 corresponding Lax pair. To determine $\overrightarrow{y}$, we use
 the following steps:

 \vspace{0.3 cm}

 {\noindent}{\bf{i)}} Find a solution $u$ of the mKdV equation
 with a given symmetry:

 \vspace{0.1 cm}
\noindent Here we consider Eq. (\ref{mKdVequation}) which is
obtained from the mKdV equation by using the traveling wave
solutions
 $u_{t}=\alpha u_{x}$, where $\alpha=-1/c$, $c\neq 0$ are arbitrary constants.

\vspace{0.3 cm}

{\noindent}{\bf{ii)}} Find the matrix $\Phi$ of the Lax equation
[Eq. (\ref{cur2})] for given $U$ and $V$:

\vspace{0.1 cm}

\noindent In our case, the corresponding ${{su}}(2)$ valued $U$
and $V$ of the mKdV equation are given by Eqs. (\ref{U mKdV}) and
(\ref{V mKdV}).
 Consider the $2\times 2$ matrix $\Phi$
 \begin{equation}{\label{phi}}
 \Phi=\left(%
\begin{array}{cc}
  \Phi_{11} & \Phi_{12} \\
  \Phi_{21} & \Phi_{22} \\
\end{array}%
\right).
 \end{equation}
\noindent By using this and Eq. (\ref{U mKdV}) for $U$, we can
write $\Phi_{x}=U\Phi$ in matrix form as
 \begin{equation}
 \left(%
\begin{array}{cc}
  (\Phi_{11})_{x} & (\Phi_{12})_{x} \\
  (\Phi_{21})_{x} & (\Phi_{22})_{x} \\
\end{array}%
\right)=\left(%
\begin{array}{cc}
 {\frac{1}{2}}\,{i\,\lambda}\,\Phi_{11}-\frac{1}{2}\,{i\,u}\,\Phi_{21} &
  {\frac{1}{2}}\,{i\,\lambda}\,\Phi_{12}-\frac{1}{2}\,{i\,u}\,\Phi_{22}\\
   -{\frac{1}{2}}\,{i\,\lambda}\,\Phi_{21}-\frac{1}{2}{i u}\,\Phi_{11} &
   -{\frac{1}{2}}\,{i\,\lambda}\,\Phi_{22}-\frac{1}{2}\,{i\,u}\,\Phi_{12} \\
\end{array}%
\right).
 \end{equation}

\noindent Combining $(\Phi_{11})_{x}= {\frac{1}{2}}\,{i
\lambda}\,\Phi_{11}-\frac{1}{2}\,{i\,u}\,\Phi_{21} $ and
$(\Phi_{21})_{x}=-{\frac{1}{2}}\,{i
\lambda}\,\Phi_{21}-\frac{1}{2}\,{i\,u}\,\Phi_{11}$, we get

 \begin{equation}{\label{Phi_21xx}}
 (\Phi_{21})_{xx}-{\frac{u_{x}}{u}}\,{(\Phi_{21})_{x}}+
 \left[\frac{1}{4\,u}\Big({u\,(\lambda^2+u^2)-2\,i\,\lambda\,u_{x}}\Big)\right]\Phi_{21}=0.
 \end{equation}

\noindent Similarly, a second order equation can be written for
$\Phi_{22}$ by using the first order equations of $\Phi_{12}$ and
$\Phi_{22}$. By solving the second order equation [Eq.
(\ref{Phi_21xx})] of $\Phi_{21}$ and the equation for $\Phi_{22}$,
we determine the explicit $x$ dependence of $\Phi_{21}, \Phi_{22}$
and also $\Phi_{11}, \Phi_{12}$. The components of
$\Phi_{t}=V\Phi$ read
 \begin{eqnarray}
&&\label{psi_11t}(\Phi_{11})_{t}=
-\frac{i}{2}\,\Big[\frac{u^2}{2}-\alpha-\alpha\,\lambda-\lambda^2\Big]\Phi_{11}-
\frac{i}{2}\,\Big[(\alpha+\lambda)\,u-i\,u_{x}\Big]\Phi_{21},\\
&&\label{psi_21t}(\Phi_{21})_{t}=
\frac{i}{2}\Big[\frac{u^2}{2}-\alpha-\alpha\lambda-\lambda^2\Big]\Phi_{21}-
\frac{i}{2}\Big[(\alpha+\lambda)u+i u_{x}\Big]\Phi_{11},
 \end{eqnarray}
and
 \begin{eqnarray}
&&\label{psi_12t}(\Phi_{12})_{t}=
-\frac{i}{2}\,\Big[\frac{u^2}{2}-\alpha-\alpha\,\lambda-\lambda^2\Big]\Phi_{12}-
\frac{i}{2}\,\Big[(\alpha+\lambda)\,u-i\,u_{x}\Big]\Phi_{22},\\
&&\label{psi_22t}(\Phi_{22})_{t}=
\frac{i}{2}\,\Big[\frac{u^2}{2}-\alpha-\alpha\,\lambda-\lambda^2\Big]\Phi_{22}-
\frac{i}{2}\,\Big[(\alpha+\lambda)\,u+i\,u_{x}\Big]\Phi_{12}.
 \end{eqnarray}
By solving these equations, we determine the explicit $t$
dependence of $\Phi_{11},$ $\Phi_{21},$ $\Phi_{12}$, and
$\Phi_{22.}$ This way we completely determine the solution $\Phi$
of the Lax equations.

\vspace{0.3 cm}

{\noindent}{\bf{iii)}} We use Eq. (\ref{eq0}) to find F. For our
case, $A$ and $B$ are given by Eqs. (\ref{A mKdV}) and (\ref{B
mKdV}), which are obtained by the spectral deformation of $U$ and
$V$, respectively. Integrating Eq. (\ref{eq0}), we get $F$.
\vspace{0.3 cm}

 Now by using a given solution of the mKdV equation, we find the
position vector of the mKdV surface. Let
$u=k_{1}\,{{\textrm{sech}}\,\xi}$,
$\displaystyle\xi=k_{1}\,\left(k_{1}^2\,t+{4\,x}\right)/8$, be one
soliton  solution of the mKdV equation, where
$\alpha={k_{1}^2}/{4}$.
 By substituting $u$ into the second order equation [Eq. (\ref{Phi_21xx})] and
 using the notation
 $u_{x}=k_{1}\,u_{\xi}/2,(\Phi_{21})_{x}=k_{1}\,(\Phi_{21})_{\xi}/2$, we find
 the solution of $\Phi_{x}=U\,\Phi$ as follows:
  \begin{eqnarray}{\label{Phi21(x)}}
  \Phi_{21}&=&iA_{1}(t)\left(\tanh{\xi}+1\right)^{i\lambda/2k_{1}}
  \left(\tanh{\xi}-1\right)^{-i\lambda/2k_{1}}{{\textrm{sech}}\,{\xi}}\\&&+
  B_{1}(t)\left(k_{1}\tanh{\xi}+2\,i\,\lambda\right)\left(\tanh{\xi}-1\right)^{i\lambda/2k_{1}}
  \left(\tanh{\xi}+1\right)^{-i\lambda/2k_{1}},\nonumber
  \end{eqnarray}
  \begin{eqnarray}{\label{Phi22(x)}}
  \Phi_{22}&=&iA_{2}(t)\left(\tanh{\xi}+1\right)^{i\lambda/2k_{1}}
  \left(\tanh{\xi}-1\right)^{-i\lambda/2k_{1}}{{\textrm{sech}}\,{\xi}}\\&&+
  B_{2}(t)\left(k_{1}\tanh{\xi}+2\,i\,\lambda\right)\left(\tanh{\xi}-1\right)^{i\lambda/2k_{1}}
  \left(\tanh{\xi}+1\right)^{-i\lambda/2k_{1}},\nonumber
  \end{eqnarray}
 \begin{eqnarray}{\label{Phi11(x)}}
 \Phi_{11}&=&-\frac{i}{k_{1}}A_{1}(t)\left(2\lambda+i\,k_{1}\tanh\,{\xi}\right)
 \left(\tanh\,{\xi}+1\right)^{i\lambda/2k_{1}}
  \left(\tanh{\xi}-1\right)^{-i\lambda/2k_{1}}\nonumber\\&&+
  i\,k_{1}\,B_{1}(t)\left(\tanh{\xi}-1\right)^{i\lambda/2k_{1}}
  \left(\tanh{\xi}+1\right)^{-i\lambda/2k_{1}}{\textrm{sech}}\,{\xi},
  \end{eqnarray}
\begin{eqnarray}{\label{Phi12(x)}}
 \Phi_{12}&=&-\frac{i}{k_{1}}A_{2}(t)\left(2\lambda+ik_{1}\tanh{\xi}\right)
 \left(\tanh{\xi}+1\right)^{i\lambda/2k_{1}}
  \left(\tanh{\xi}-1\right)^{-i\lambda/2k_{1}}\nonumber\\&&+
  i\,k_{1}\,B_{2}(t)\left(\tanh{\xi}-1\right)^{i\lambda/2k_{1}}
  \left(\tanh{\xi}+1\right)^{-i\lambda/2k_{1}}{\textrm{sech}}\,{\xi}.
  \end{eqnarray}
Hence one part $(\Phi_{x}=U\Phi)$ of the Lax equations has been
solved. By using these solutions in Eqs.
(\ref{psi_11t})-(\ref{psi_22t}) obtained from $\Phi_{t}=V\Phi$, we
find
\begin{equation}
A_{1}(t)=A_{1}e^{i\left(k_{1}^2+4\lambda^2\right)t/8}~~
{\textrm{and}}
~~B_{1}(t)=B_{1}e^{-i\left(k_{1}^2+4\lambda^2\right)t/8},
 \end{equation}
\begin{equation}
A_{2}(t)=A_{2}e^{i\left(k_{1}^2+4\lambda^2\right)t/8}~~
{\textrm{and}}
~~B_{2}(t)=B_{2}e^{-i\left(k_{1}^2+4\lambda^2\right)t/8},
 \end{equation}
where $A_{1},$ $A_{2}$, $B_{1}$, and $B_{2}$ are arbitrary
constants. We solved the Lax equations for a given $U,$ $V$ and a
solution $u$ of the mKdV equation [Eq. (\ref{mKdVequation})]. The
components of $\Phi$ are
\begin{eqnarray}{\label{Phii11(x)}}
 \Phi_{11}&=&-\frac{i}{k_{1}}{A_{1}\,e^{i\left(k_{1}^2+4\lambda^2\right)t/8}}
 \left(2\,\lambda+i\,k_{1}\,\tanh{\xi}\right)
 \left(\tanh{\xi}+1\right)^{i\lambda/2k_{1}}
  \left(\tanh{\xi}-1\right)^{-i\lambda/2k_{1}}\nonumber\\&&+
  i\,k_{1}\,{B_{1}\,e^{-i\left(k_{1}^2+4\lambda^2\right)t/8}}\left(\tanh{\xi}-1\right)^{i\lambda/2k_{1}}
  \left(\tanh{\xi}+1\right)^{-i\lambda/2k_{1}}{\textrm{sech}}\,{\xi},
  \end{eqnarray}
\begin{eqnarray}{\label{Phii12(x)}}
 \Phi_{12}&=&-\frac{i}{k_{1}}{A_{2}\,e^{i\left(k_{1}^2+4\lambda^2\right)t/8}}
 \left(2\,\lambda+i\,k_{1}\,\tanh{\xi}\right)
 \left(\tanh{\xi}+1\right)^{i\lambda/2k_{1}}
  \left(\tanh{\xi}-1\right)^{-i\lambda/2k_{1}}\nonumber\\&&+
  i\,k_{1}\,{B_{2}\,e^{-i\left(k_{1}^2+4\lambda^2\right)t/8}}\left(\tanh{\xi}-1\right)^{i\lambda/2k_{1}}
  \left(\tanh{\xi}+1\right)^{-i\lambda/2k_{1}}{\textrm{sech}}\,{\xi}.
  \end{eqnarray}
  \begin{eqnarray}{\label{Phii21(x)}}
  \Phi_{21}&=&i\,{A_{1}\,e^{i\left(k_{1}^2+4\lambda^2\right)t/8}}\left(\tanh{\xi}+1\right)^{i\lambda/2k_{1}}
  \left(\tanh{\xi}-1\right)^{-i\lambda/2k_{1}}{{\textrm{sech}}\,{\,\xi}}\\&&+
  {B_{1}\,e^{-i\left(k_{1}^2+4\lambda^2\right)t/8}}
  \left(k_{1}\tanh{\xi}+2i\lambda\right)\left(\tanh{\xi}-1\right)^{i\lambda/2k_{1}}
  \left(\tanh{\xi}+1\right)^{-i\lambda/2k_{1}}\nonumber,
  \end{eqnarray}
  \begin{eqnarray}{\label{Phii22(x)}}
  \Phi_{22}&=&i\,{A_{2}\,e^{i\left(k_{1}^2+4\lambda^2\right)t/8}}\left(\tanh{\xi}+1\right)^{i\lambda/2k_{1}}
  \left(\tanh{\xi}-1\right)^{-i\lambda/2k_{1}}{{\textrm{sech}}\,{\,\xi}}\\&&+
 {B_{2}\,e^{-i\left(k_{1}^2+4\lambda^2\right)t/8}}
 \left(k_{1}\tanh{\xi}+2i\lambda\right)\left(\tanh{\xi}-1\right)^{i\lambda/2k_{1}}
  \left(\tanh{\xi}+1\right)^{-i\lambda/2k_{1}}.\nonumber
  \end{eqnarray}

\noindent Here we find that
$\det(\Phi)=[{(k_{1}^2+4\lambda^2)}/{k_{1}}]\left(A_{1}B_{2}-A_{2}B_{1}\right)\neq
0$.

Inserting $A, B$, and $\Phi$ in Eq. (\ref{eq0}), and solving the
resultant equation and letting $A_{1}=A_{2}$,
$B_{1}=\displaystyle\left(A_{1}\,{e^{\pi\lambda/k_{1}}}\right)/k_{1},$
$B_{2}=-B_{1}$, we obtain a three parameter ($\lambda, k_{1},
\mu$) family of surfaces parametrized by
\begin{eqnarray}
&&\label{y1}y_{1}=-\frac{1}{4\,k_{1}\,(e^{2\xi}+1)}R_{1}\left(E(e^{2\xi}+1)+32k_{1}\right),\\
&&\label{y2}y_{2}=-{4\,R_{1}\,\cos{G}}\,{{\textrm{sech}}\,{\xi}},\\
&&\label{y3}y_{3}={4\,R_{1}\,\sin{G}}}\,{{\textrm{sech}}\,{\xi},
\end{eqnarray}
where
\begin{eqnarray}
&&R_{1}=\displaystyle-\frac{\mu\,k_{1}}{2\,(k_{1}^2+4\,\lambda^2)},\\
&&G=\displaystyle
t\left(\lambda^2+\frac{1}{4}\,k_{1}^2[1+\lambda]\right)+x\,\lambda,\\
&&E=\left(t\,[8\,\lambda+k_{1}^2]+4\,x\right)\left(k_{1}^2+4
\,\lambda^2\right),\\
&&\xi=\frac{k_{1}^3}{8}(t+\frac{4x}{k_{1}^2}).
\end{eqnarray}
 This surface
 has the following first and second fundamental forms:
 \begin{eqnarray}
 &&(ds_{I})^2=\frac{1}{4}\,{\mu^2}\,
 \Bigg[\Big(dx+\big[\frac{1}{4}\,k_{1}^2+2\lambda\big]dt\Big)^2
 +k_{1}^2\,{\textrm{sech}}^2\,{\xi}\,dt^2\Bigg],\\
&&(ds_{II})^2=\frac{1}{2}{\mu}\,{k_{1}\,{\textrm{sech}}\,{\xi}}
\left[dx+\left(\frac{1}{4}\,k_{1}^2+\lambda\right)dt\right]^2
+\frac{1}{8}\,{\mu}\,{k_{1}^3\,{\textrm{sech}}\,{\xi}}
\left[2\,{\textrm{sech}}^2\,{\xi}-1\right]dt^2\nonumber.
 \end{eqnarray}
and the Gaussian and mean curvatures, respectively, are
 \begin{eqnarray}
 &&K=\frac{k_{1}^2}{\mu^2}\Big(2\,{\textrm{sech}}^2\,{\xi}-1\Big),{\label{Weingarten K}}\\
 &&H=\frac{1}{4\,\mu\,k_{1}\,{{\textrm{sech}}\,{\xi}}}
 \Big(6\,k_{1}^2\,{\textrm{sech}}^2\,{\xi}+\,(4\,\lambda^2-k_{1}^2)\Big).{\label{Weingarten H}}
 \end{eqnarray}

\vspace{0.3 cm}

 \noindent  {\it{{Proposition 4:}}\,The surface which is parametrized
 by Eqs.
 (\ref{y1})-(\ref{y3}) is a cubic
 Weingarten surface, i.e.,
 \begin{equation}
 4\,\mu^2\,H^2\left(2[\mu^2\,K+k_{1}^2]\right)-9\,\mu^4\,K^2-12\,\mu^2\,\left(k_{1}^2+2\,\lambda^2\right)\,K
 -\left(k_{1}^2+2\,\lambda^2\right)^2=0.
 \end{equation}
 When $k_{1}=2\,\lambda$ in Eqs. (\ref{Weingarten K}) and (\ref{Weingarten H}), it reduces to a quadratic Weingarten
 surface, i.e.,
 \begin{equation}
 8\,\mu^2\,H^2-9\,\mu^2\,K-36\,\lambda^2=0.
 \end{equation}
}

\vspace{0.3 cm}

\subsection{The analyses of the three parameter family of mKdV surfaces}

\vspace{0.3 cm}

 In general, $y_{2}$ and $y_{3}$ are asymptotically decaying
 functions,
  and $y_{1}$ approaches $\pm\infty$ as $\xi$ tends to
$\pm\infty$. For some small intervals of $x$ and $t$, we plot some
of the three parameter family of surfaces for some special values
of the parameters $k_{1},$ $\lambda$, and $\mu$ in Figs.
{{{\ref{fig1}}}-{{{\ref{fig4}}}.

\vspace{0.3 cm}

 \noindent{\it{Example 2:}}{\label{simplified mkdv
surface}} By taking $k_{1}=2,$ $\lambda=1,$ and $\mu=-8$ in Eqs.
(\ref{y1})-(\ref{y3}), we get the surface (Fig. \ref{fig1}).

\begin{figure}[!h]
\centering
{\includegraphics[height=5cm,width=5cm,scale=.28]{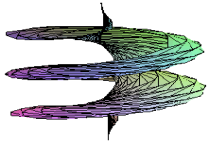}}
\caption{$(x,t)\in[-3,3]\times[-3,3]$}\label{fig1}
\end{figure}

The components of the position vector of the surface are
\begin{equation}
y_{1}=-{{E_{1}}}-8/({e^{2\xi}+1}), y_{2}=-4\cos{G}\,
{\textrm{sech}}\,{\xi},y_{3}=-4\sin{G}\, {\textrm{sech}}\,{\xi},
\end{equation}
where ${E_{1}}=4(x+3t),$ $G=x+3t$, and $\xi=x+t.$ As $\xi$ tends
to $\pm\infty,$ $y_{1}$ approaches $\pm\infty,$ and $y_{2}$ and
$y_{3}$ approach zero. This can also be seen in Fig. \ref{fig1}.
For small values of $x$ and $t$, the surface has a twisted shape.

\vspace{0.3 cm}

\noindent{\it{Example 3:}} By taking $k_{1}=2,$ $\lambda=0,$ and
$\mu=-4$ in Eqs. (\ref{y1})-(\ref{y3}), we get the surface (Fig.
\ref{fig2}).

\begin{figure}[!h]
\centering
{\includegraphics[height=5cm,width=5cm,scale=.28]{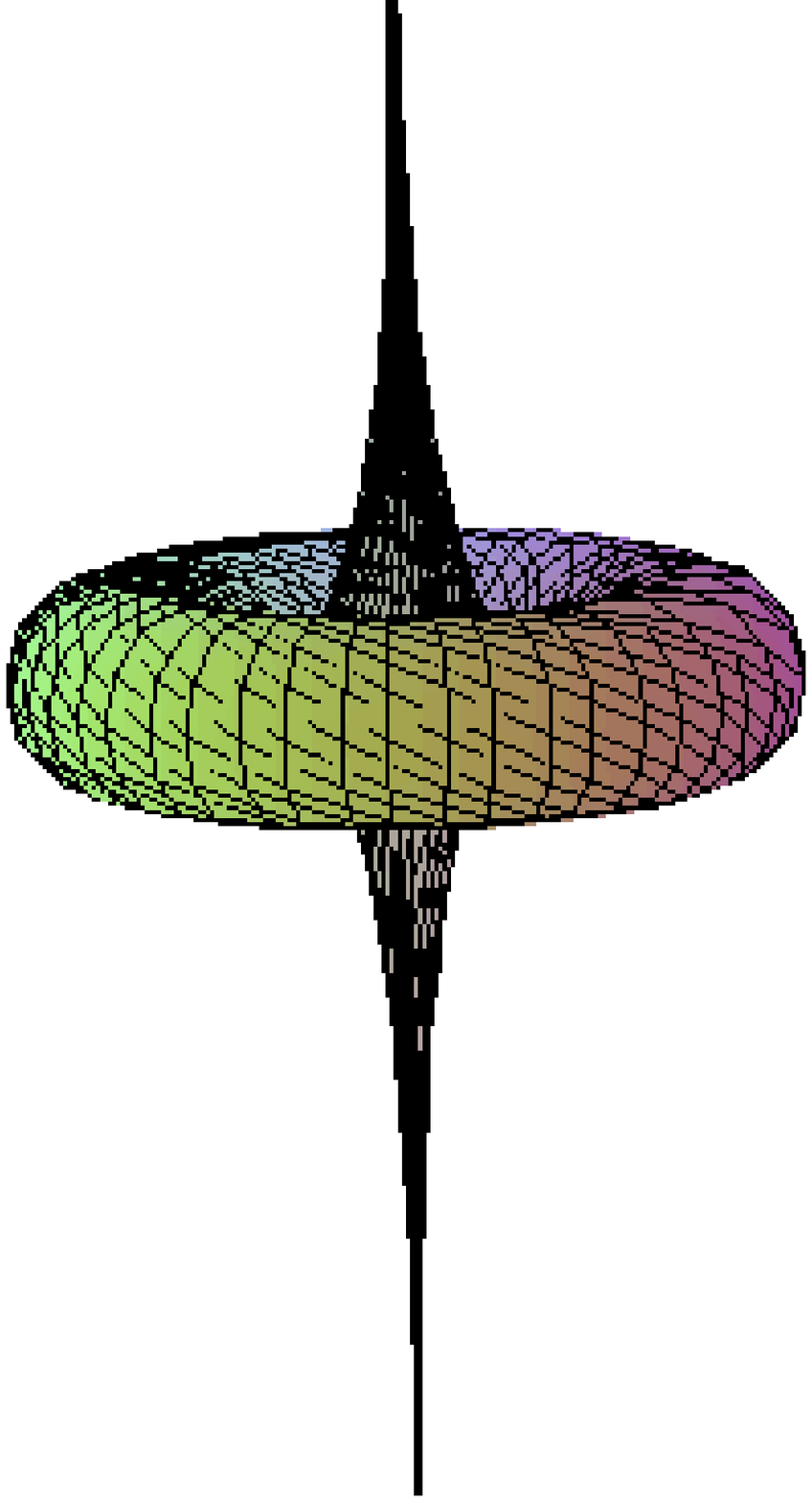}}
\caption{$(x,t)\in[-6,6]\times[-6,6]$}\label{fig2}
\end{figure}

The components of the position vector of the surface are
\begin{equation}
y_{1}=-{{E_{1}}}-8/(e^{2\xi}+1), y_{2}=-4\cos{G}\,
{\textrm{sech}}\,{\xi}, y_{3}=-4\sin{G}\, {\textrm{sech}}\,{\xi},
\end{equation}
where ${E_{1}}=2(x+3t),$ $G=t$, and $\xi=x+t.$ As $\xi$ tends to
$\pm\infty,$ $y_{1}$ approaches $\pm\infty,$ and $y_{2}$ and
$y_{3}$ tend to zero. This can also be seen in Fig. \ref{fig2}.
Asymptotically, this surface and the surface given in Example 2
are the same. However, for small values of $x$ and $t$, they are
different.

\vspace{0.3 cm}

 \noindent{\it{Example 4:}} By taking $k_{1}=3,$
$\lambda={1}/{10},$ and $\mu=-{452}/{75}$ in Eqs.
(\ref{y1})-(\ref{y3}), we get the surface (Fig. \ref{fig3}).

\begin{figure}[!h]
\centering
{\includegraphics[height=5cm,width=5cm,scale=.28]{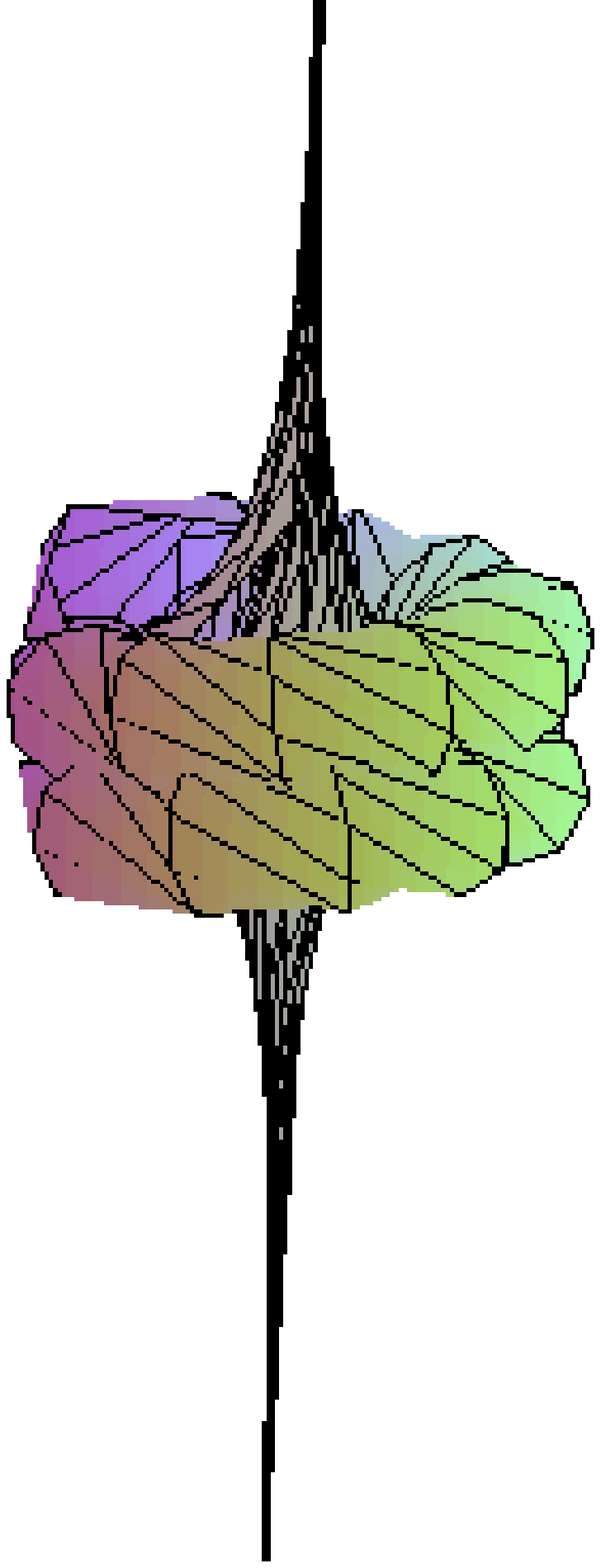}}
\caption{$(x,t)\in[-6,6]\times[-6,6]$}\label{fig3}
\end{figure}

The components of the position vector of the surface are
\begin{equation}
y_{1}=-{{E_{1}}}-8/(e^{2\xi}+1), y_{2}=-4\cos{G}\,
{\textrm{sech}}\,{\xi}, y_{3}=-4\sin{G}\, {\textrm{sech}}\,{\xi},
\end{equation}
where ${E_{1}}=-(5537\,t+2260\,x)/750,$
$G=17\,(497\,t+20\,x)/200$, and $\xi=(12\,x+27\,t)/8.$
Asymptotically, this surface is similar to the previous two
surfaces. For small values of $x$ and $t$, the surface looks like
a shell.

\vspace{0.3 cm}

\noindent{\it{Example 5:}} By taking $k_{1}=1,$
$\lambda=-{1}/{10},$ and $\mu=-{52}/{25}$ in Eqs.
(\ref{y1})-(\ref{y3}), we get the surface (Fig. \ref{fig4}).

\begin{figure}[!h]
\centering
{\includegraphics[height=5cm,width=5cm,scale=.28]{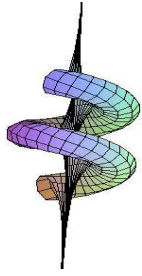}}
\caption{$(x,t)\in[-20,20]\times[-20,20]$}\label{fig4}
\end{figure}

The components of the position vector of the surface are
\begin{equation}
y_{1}=-{{E_{1}}}-8/(e^{2\xi}+1), y_{2}=-4\cos{G}\,
{\textrm{sech}}\,{\xi}, y_{3}=-4\sin{G}\, {\textrm{sech}}\,{\xi},
\end{equation}
where ${E_{1}}=-13(20\,x+t)/250,$ $G=(47\,t-20\,x)/200$, and
$\xi=(4\,x+t)/8.$ Asymptotically, this surface is similar to the
previous three surfaces.

\vspace{0.3 cm}

\section{mKdV Surfaces From Spectral-Gauge Deformations}

 In this section, we find surfaces arising from a combination of the
 spectral and gauge deformations of the Lax pair for the mKdV equation.

\vspace{0.3 cm}

\noindent  {\it{{Proposition 5:}}{\label{mKdVSpecGau}}
 \,Let $u$ satisfy (which describes a traveling mKdV wave)
 Eq. (\ref{mKdVequation}). The corresponding ${{su}}(2)$ valued Lax
pair $U$ and $V$ of the mKdV equation are given by Eqs. (\ref{U
mKdV}) and (\ref{V mKdV}), respectively. The ${{su}}(2)$ valued
matrices $A$ and $B$ are
   \begin{eqnarray}{\label{ASpecGau mKdV}}
         A=i\left(%
              \begin{array}{cc}
                  (\frac{1}{2}\,\mu-\nu\,u) &  -\nu \lambda \\
                  -\nu \lambda & -(\frac{1}{2}\,\mu-\nu\,u)  \\
               \end{array}%
                \right),
                \end{eqnarray}
             \begin{eqnarray}{\label{BSpecGau mKdV}}
          B={i}\left(%
          \begin{array}{cc}
          \frac{1}{2}\,\mu\,(\alpha+2\,\lambda)-\nu\,(\alpha+\lambda)\,u &
           -\frac{1}{2}\,\mu\,u+\nu\,(\frac{1}{2}\,u^2-\alpha-\alpha\,\lambda-\lambda^2) \\
           -\frac{1}{2}\,\mu \,u+\nu\,(\frac{1}{2}\,u^2-\alpha-\alpha\,\lambda-\lambda^2) &
           -\frac{1}{2}\,\mu\,(\alpha+2\lambda)+\nu\,(\alpha+\lambda)\,u \\
          \end{array}%
                \right),
   \end{eqnarray}
  where $A=\mu\,U_{\lambda}+\nu\,[\sigma_{2},U],$
  $B=\mu\,V_{\lambda}+\nu\,[\sigma_{2},V]$, $\lambda$ is the spectral parameter,
  $\mu$ and $\nu$ are constants, and $\sigma_{2}$ is the Pauli sigma matrix.
  The surface $S$, generated by $U,V,A$, and $B$, has the following first and
  second fundamental forms ($j,k=1,2$)
 \begin{eqnarray}
       &&(ds_{I})^2=g_{jk}\,dx^{j}\,dx^{k},\\
        &&(ds_{II})^2=h_{jk}\,dx^{j}\,dx^{k},
 \end{eqnarray}
 where
  \begin{eqnarray}
       &&g_{11}=\frac{1}{4}\,\mu^2+\nu\,(\nu\,[u^2+\lambda^2]-\mu\,u),\\
       &&g_{12}=\frac{1}{4}(\alpha+2\,\lambda)\mu^2+
       \frac{1}{4}\nu\Big(\nu\big[{2}(\lambda+2\,\alpha)u^2+4\,(\lambda^3
       +\alpha\,\lambda+\lambda^2\,\alpha)\big]\nonumber\\
       &&~~~~~~~~-4\,\mu\,(\alpha+\lambda)\,u\Big),\\
       &&g_{22}=\frac{1}{4}(u^2+(2\,\lambda+\alpha)^2)\mu^2+
       \nu\,\bigg(\nu\,\big[\frac{1}{4}\,u^4+\alpha\,(\alpha-1+\lambda)u^2\nonumber\\
       &&~~~~~~~+((1+\lambda)\alpha+\lambda^2)^2\big]
       -\frac{1}{2}\,\mu\,
       u^3-\mu\,(\alpha^2+(2\,\lambda-1)\,\alpha+\lambda^2)\,u\bigg),\\
       &&h_{11}=\frac{1}{2}\,\mu\,u-\nu(u^2+\lambda^2),\\
       &&h_{12}=\frac{1}{2}\,\mu\,(\alpha+\lambda)\,u-
       \nu\,\Big(\lambda(\lambda^2+\alpha\,\lambda+\alpha)+\frac{1}{2}\,(\lambda+2\,\alpha)u^2\Big),\\
       &&h_{22}=\frac{1}{4}\,\mu\,\Big(u^3+2\big[\alpha^2+(2\,\lambda-1)\alpha+\lambda^2\big]\,u\Big)\nonumber\\
       &&~~~~~~~-\nu\,\Big(\frac{1}{4}\,u^4+
       \alpha(\alpha-1+\lambda)u^2+((1+\lambda)\,\alpha+\lambda^2)^2\Big),
 \end{eqnarray}
 and the corresponding Gaussian and mean curvatures are
 \begin{eqnarray}
 &&K=\frac{2\,u\,(u^2-2\,\alpha)}{\nu\Big(2\,\nu u[u^2-2\,\alpha]-3\,\mu
 u^2-2\mu(\lambda^2-\alpha)\Big)+\mu^2 u},\\
 &&H=\frac{\mu(3\,u^2+2(\lambda^2-\alpha))-4\,u\,\nu(u^2-2\,\alpha)}{2\,\nu\Big(2\,\nu\,u[u^2-2\,\alpha]
 -3\,\mu\,u^2-2\,\mu(\lambda^2-\alpha)\Big)+2\,\mu^2\,u},
 \end{eqnarray}
 where $x^1=x$ and $x^2=t.$
 }

\vspace{0.3 cm}

\subsection{The parametrized form of the four parameter family of mKdV surfaces}

\vspace{0.3 cm}

We apply the same technique that we used in section 2 to find the
position vector of the corresponding surface. Let
$u=k_{1}\,{{\textrm{sech}}\,\xi}$,
$\displaystyle\xi=k_{1}\,\left(k_{1}^2\,t+{4x}\right)/8$, be the
one soliton solution of the mKdV equation, where
$\alpha=\displaystyle{k_{1}^2}/{4}$. The Lax pair $U$ and $V$ are
given by Eqs. (\ref{U mKdV}) and (\ref{V mKdV}), respectively,
which is the same as in the spectral deformation case. So we can
use the solution of the Lax equation [Eq. (\ref{cur2})] that we
found in the spectral deformation case. By solving Eq.
(\ref{eq0}), we obtain the position vector, where the components
of $\Phi$ are given by Eqs. (\ref{Phii11(x)})-(\ref{Phii22(x)})
and $A$, $B$ are given by Eqs. (\ref{ASpecGau mKdV}) and
(\ref{BSpecGau mKdV}), respectively. Here we choose $A_{1}=A_{2},$
$B_{1}=\displaystyle\left(A_{1}\,{e^{\pi\lambda/k_{1}}}\right)/{k_{1}},$
$B_{2}=-B_{1}$ to write $F$ in the form
$F=i(\sigma_{1}y_{1}+\sigma_{2}y_{2}+\sigma_{3}y_{3})$. Hence we
obtain a four parameter ($\lambda, k_{1}, \mu, \nu$) family of
surfaces parametrized by
\begin{eqnarray}
&&\label{y1SpecGau}y_{1}=R_{2}\,{\frac{e^{2\xi}-1}{(e^{2\xi}+1)}}\,{\textrm{sech}}\,{\xi}
+R_{3}\,\tilde{E}+R_{4}\frac{1}{e^{2\xi}+1},\\
&&\label{y2SpecGau}y_{2}=\Big[\frac{1}{2}\,R_{4}\,{\textrm{sech}}\,{\xi}+
R_{5}\,{\frac{\big(e^{4\xi}+1\big)}{(e^{2\xi}+1)^2}}-R_{6}\,{\textrm{sech}}^2\,{\xi}\Big]\cos{G}
+R_{7}\,\frac{\big(e^{2\xi}-1\big)}{\big(e^{2\xi}+1\big)}\sin{G},\nonumber\\
&&\label{y3SpecGau}y_{3}=\Big[-\frac{1}{2}\,R_{4}\,{\textrm{sech}}\,{\xi}-
R_{5}\,{\frac{\big(e^{4\xi}+1\big)}{(e^{2\xi}+1)^2}}+R_{6}\,{\textrm{sech}}^2\,{\xi}\Big]\sin{G}
+R_{7}\,\frac{\big(e^{2\xi}-1\big)}{\big(e^{2\xi}+1\big)}\cos{G},\nonumber
\end{eqnarray}
where
\begin{eqnarray}
&&R_{2}=\frac{2\,k_{1}^2\,\nu}{k_{1}^2+4\lambda^2},~~R_{3}=\frac{\mu}{8},\\
&&R_{4}=\frac{4\,\mu\,k_{1}^2}{k_{1}^2+4\lambda^2},~~
R_{5}=\frac{\nu\,(k_{1}^2-4\lambda^2)}{k_{1}^2+4\lambda^2},\\
&&R_{6}=\frac{\nu\,(4\,\lambda^2+3\,k_{1}^2)}{2(k_{1}^2+4\lambda^2)},~~
R_{7}=\frac{4\,\lambda\,k_{1}^2\,\nu}{k_{1}^2+4\lambda^2},
\end{eqnarray}
\begin{eqnarray}
&&G=\displaystyle
t\left(\lambda^2+\frac{1}{4}\,k_{1}^2\,[1+\lambda]\right)+x\lambda,\\
&&\tilde{E}=\left(t\,[8\,\lambda+k_{1}^2]+4\,x\right),\\
&&\xi=\frac{k_{1}^3}{8}(t+\frac{4\,x}{k_{1}^2}).
\end{eqnarray}
 Thus the position vector  $\overrightarrow{y}=(y_{1}(x,t),y_{2}(x,t),y_{3}(x,t))$
 of the surface is given by Eq. (\ref{y1SpecGau}). This surface
 has the following first and second fundamental forms:
\begin{eqnarray}
       &&(ds_{I})^2=g_{jk}\,dx^{j}\,dx^{k},\\
        &&(ds_{II})^2=h_{jk}\,dx^{j}\,dx^{k},
 \end{eqnarray}
 where
\begin{eqnarray}
       &&g_{11}=\frac{1}{4}\,\mu^2+
       \nu\,(\nu\,[k_{1}^2 \,{\textrm{sech}}^2\,{\xi}+\lambda^2]
       -\mu\,{k_{1}\,{\textrm{sech}}\,{\xi}}),\nonumber\\
       &&g_{12}=\frac{1}{4}(\alpha+2\,\lambda){\mu^2}+
       \frac{1}{4}\,\nu\,\Big(\nu\,\big[{2}\,k_{1}^2(\lambda+2\,\alpha)\,{\textrm{sech}}^2\,{\xi}
       +(4\,\lambda^3+4\,\alpha\,\lambda+4\,\lambda^2\,\alpha)\big]\nonumber\\
       &&~~~~~~~-4\,\mu\,(\alpha+\lambda)\,{k_{1}\,{\textrm{sech}}\,{\xi}}\Big),\nonumber
              \end{eqnarray}
       \begin{eqnarray}
       &&g_{22}=\frac{1}{4}\,(k_{1}^2\,{\textrm{sech}}^2\,{\xi}+(2\,\lambda+\alpha)^2)\mu^2+
       \nu\,\bigg(\nu\,\Big[\frac{1}{4}\,k_{1}^4\,{\textrm{sech}}^4\,{\xi}
       +\alpha\,k_{1}^2(\alpha-1+\lambda)\,{\textrm{sech}}^2\,{\xi}\nonumber\\
       &&~~~~~~~+((1+\lambda)\alpha+\lambda^2)^2\Big]
       -\frac{1}{2}\,\mu
       \,k_{1}^3\,{\textrm{sech}}^3\,{\xi}-\mu\,{k_{1}\,(\alpha^2
       +(2\,\lambda-1)\alpha+\lambda^2)\,{\textrm{sech}}\,{\xi}}\bigg),\nonumber\\
       &&h_{11}=\frac{1}{2}\,\mu\,{k_{1}\,{\textrm{sech}}\,{\xi}}-\nu\,(k_{1}^2 \,{\textrm{sech}}^2\,{\xi}
       +\lambda^2),\nonumber\\
       &&h_{12}=\frac{1}{2}\,\mu(\alpha+\lambda)\,{k_{1}\,{\textrm{sech}}\,{\xi}}-
       \nu\,\Big(\lambda(\lambda^2+\alpha\,\lambda+\alpha)
       +\frac{1}{2}\,k_{1}^2\,(\lambda+2\,\alpha)\,{\textrm{sech}}^2\,{\xi}\Big),\nonumber\\
       &&h_{22}=\frac{1}{4}\,\mu(k_{1}^3 \,{\textrm{sech}}^3\,{\xi}
       +2\,k_{1}^2 \,[\alpha^2+(2\,\lambda-1)\alpha+\lambda^2]\,{\textrm{sech}}^2\,{\xi})\nonumber\\
       &&~~~~~~~-\nu\,\Big(\frac{1}{4}\,k_{1}^4 \,{\textrm{sech}}^4\,{\xi}+
       \alpha\,k_{1}^2\,(\alpha-1+\lambda) \,{\textrm{sech}}^2\,{\xi}+((1+\lambda)\alpha+\lambda^2)^2\Big),\nonumber
 \end{eqnarray}
and the corresponding Gaussian and mean curvatures are
 \begin{eqnarray}
 &&K=\frac{2\,k_{1}\,{\textrm{sech}}\,{\xi}(k_{1}^2\,{\textrm{sech}}^2\,{\xi}-2\,\alpha)}
 {\nu\,\Big(2\,\nu\,k_{1}\,{\textrm{sech}}\,{\xi}[k_{1}^2\,{\textrm{sech}}^2\,{\xi}-2\,\alpha]-3\,\mu\,
 k_{1}^2\,{\textrm{sech}}^2\,{\xi}-2\,\mu\,(\lambda^2-\alpha)\Big)+\mu^2\,k_{1}\,{\textrm{sech}}\,{\xi}},\nonumber
 \end{eqnarray}
\begin{eqnarray}
 &&H=\frac{\mu\,(3\,k_{1}^2\,{\textrm{sech}}^2\,{\xi}+2\,(\lambda^2-\alpha))
 -4\,\nu\,k_{1}\,{\textrm{sech}}\,{\xi}(k_{1}^2\,{\textrm{sech}}^2\,{\xi}-2\,\alpha)}
 {2\,\nu\,\Big(2\,\nu\,k_{1}\,{\textrm{sech}}\,{\xi}[k_{1}^2\,{\textrm{sech}}^2\,{\xi}-2\,\alpha]
 -3\,\mu\,k_{1}^2\,{\textrm{sech}}^2\,{\xi}-2\,\mu(\lambda^2-\alpha)\Big)
 +2\,\mu^2\,k_{1}\,{\textrm{sech}}\,{\xi}},\nonumber
 \end{eqnarray}
 where $x^1=x,$ $x^2=t$, and $\alpha=\frac{1}{4}k_{1}^2$.

\vspace{0.3 cm}

\subsection{The analyses of the four parameter family of surfaces}

\vspace{0.3 cm}

Asymptotically, $y_{1}$ approaches $\pm\infty$, $y_{2}$ approaches
$R_{5}\,\cos{G}\pm R_{7}\,\sin{G}$, and $y_{3}$ approaches
$-R_{5}\,\sin{G}\pm R_{7}\,\cos{G}$ as $\xi$ tends to $\pm\infty$.
For some small intervals of $x$ and $t$, we plot some of the four
parameter family of surfaces for some special values of the
parameters $k_{1},$ $\lambda,$ $\mu$, and $\nu$ in Figs.
{\ref{fig5}}-{\ref{fig7}}.

\vspace{0.3 cm}

\noindent{\it{Example 6:}} By taking $k_{1}=2,\lambda=0, \mu=-4,$
and $\nu=1$ in Eq. (\ref{y1SpecGau}), we get the surface (Fig.
\ref{fig5}).

\begin{figure}[!h]
\centering
{\includegraphics[height=5cm,width=5cm,scale=.28]{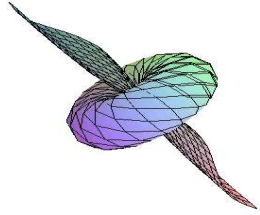}}
\caption{$(x,t)\in[-4,4]\times[-4,4]$}\label{fig5}
\end{figure}

The components of the position vector are
\begin{eqnarray}
&&y_{1}=2\,{\textrm{sech}}\,{\xi}\,{{(e^{2\xi}-1)}/{(e^{2\xi}+1)}}
-{{E_{2}}}-8/({e^{2\xi}+1}),\\
&&y_{2}=\big[-4\,{\textrm{sech}}\,{\xi}+
\,{{\big(e^{4\xi}+1\big)}/{(e^{2\xi}+1)^2}}-
(3/2)\,{\textrm{sech}}^2\,{\xi}\big]\cos{G},\\
&&y_{3}=\big[4\,{\textrm{sech}}\,{\xi}-
{{\big(e^{4\xi}+1\big)}/{(e^{2\xi}+1)^2}}+
(3/2)\,{\textrm{sech}}^2\,{\xi}\big]\sin{G},
\end{eqnarray}
 where
${{E_{2}}}=-2(x+t),$ $G=t$, and $\xi=x+t.$ As $\xi$ tends to
$\pm\infty$, $y_{1}$ approaches $\pm\infty$, $y_{2}$ approaches
$\cos{t}$, and $y_{3}$ approaches $-\sin{t}.$

\vspace{0.3 cm}

\noindent{\it{Example 7:}} By taking $k_{1}=2,\lambda=1,
\mu=1/10,$ and $\nu=1$ in Eq. (\ref{y1SpecGau}), we get the
surface (Fig. \ref{fig6}).

\begin{figure}[!h]
\centering
{\includegraphics[height=5cm,width=5cm,scale=.28]{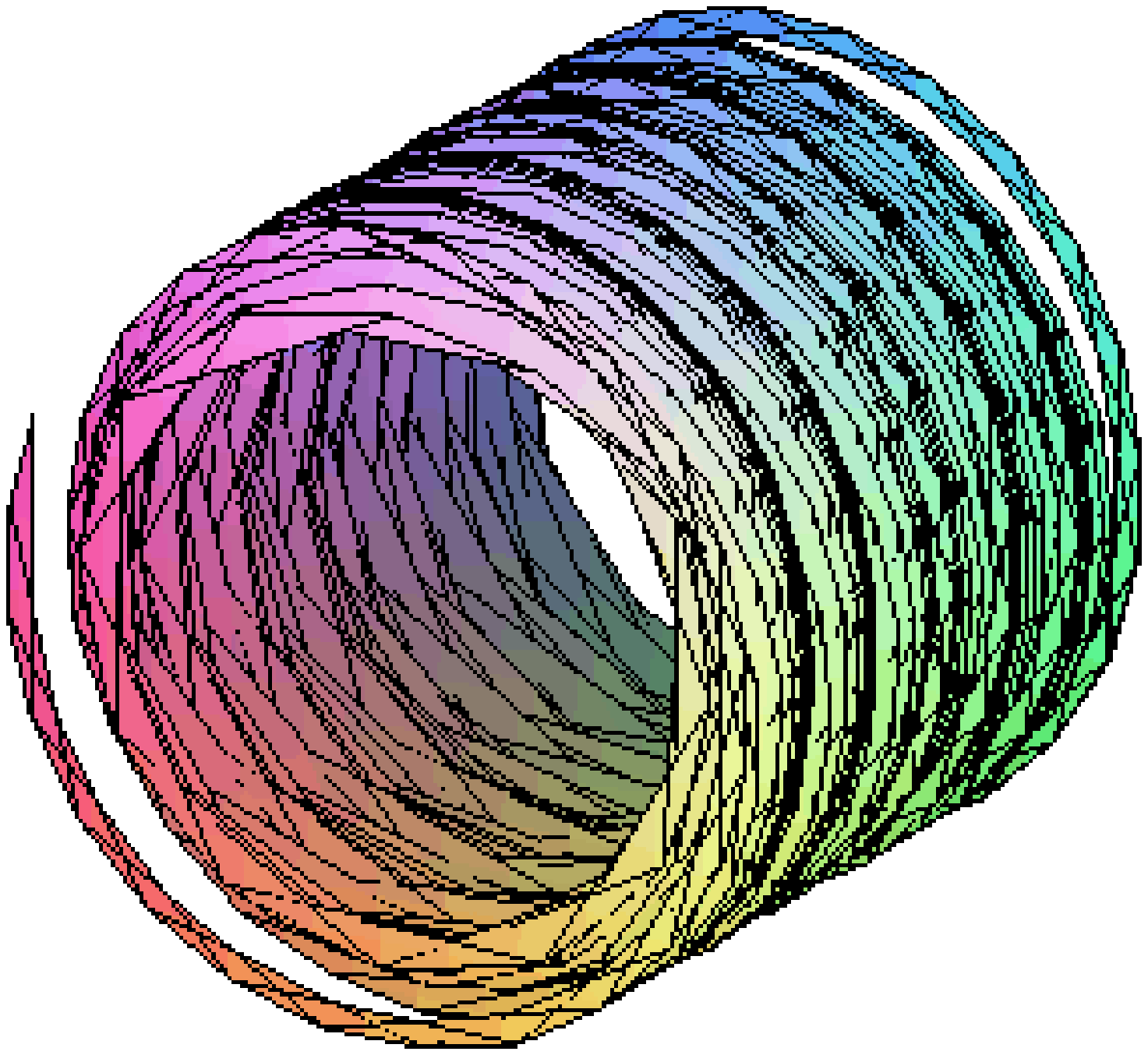}}
\caption{$(x,t)\in[-6,6]\times[-6,6]$}\label{fig6}
\end{figure}

The components of the position vector are
\begin{eqnarray}
&&y_{1}={\textrm{sech}}\,{\xi}\,{{(e^{2\xi}-1)}/{(e^{2\xi}+1)}}
+{{E_{2}}}+1/(10({e^{2\xi}+1})),\\
&&y_{2}=\big[(1/20)\,{\textrm{sech}}\,{\xi}-{\textrm{sech}}^2\,{\xi}\big]\cos{G}
+\sin{G}\,{\big(e^{2\xi}-1\big)}/{\big(e^{2\xi}+1\big)},\\
&&y_{3}=\big[-(1/20)\,{\textrm{sech}}\,{\xi}+{\textrm{sech}}^2\,{\xi}\big]\sin{G}
+\cos{G}\,{\big(e^{2\xi}-1\big)}/{\big(e^{2\xi}+1\big)},
\end{eqnarray}
 where
${{E_{2}}}=(x+3t)/20,$ $G=(x+3t)$, and $\xi=x+t.$ As $\xi$ tends
to $\pm\infty$, $y_{1}$ tends to $\pm\infty$, $y_{2}$ approaches
$\sin{(x+3t)}$, and $y_{3}$ approaches $\cos{(x+3t)}.$

\vspace{0.3 cm}

\noindent{\it{Example 8:}} By taking $k_{1}=1,\lambda=-1/10,
\mu=-52/25,$ and $\nu=-1$ in Eq. (\ref{y1SpecGau}), we get the
surface (Fig. \ref{fig7}).

\begin{figure}[!h]
\centering
{\includegraphics[height=5cm,width=5cm,scale=.28]{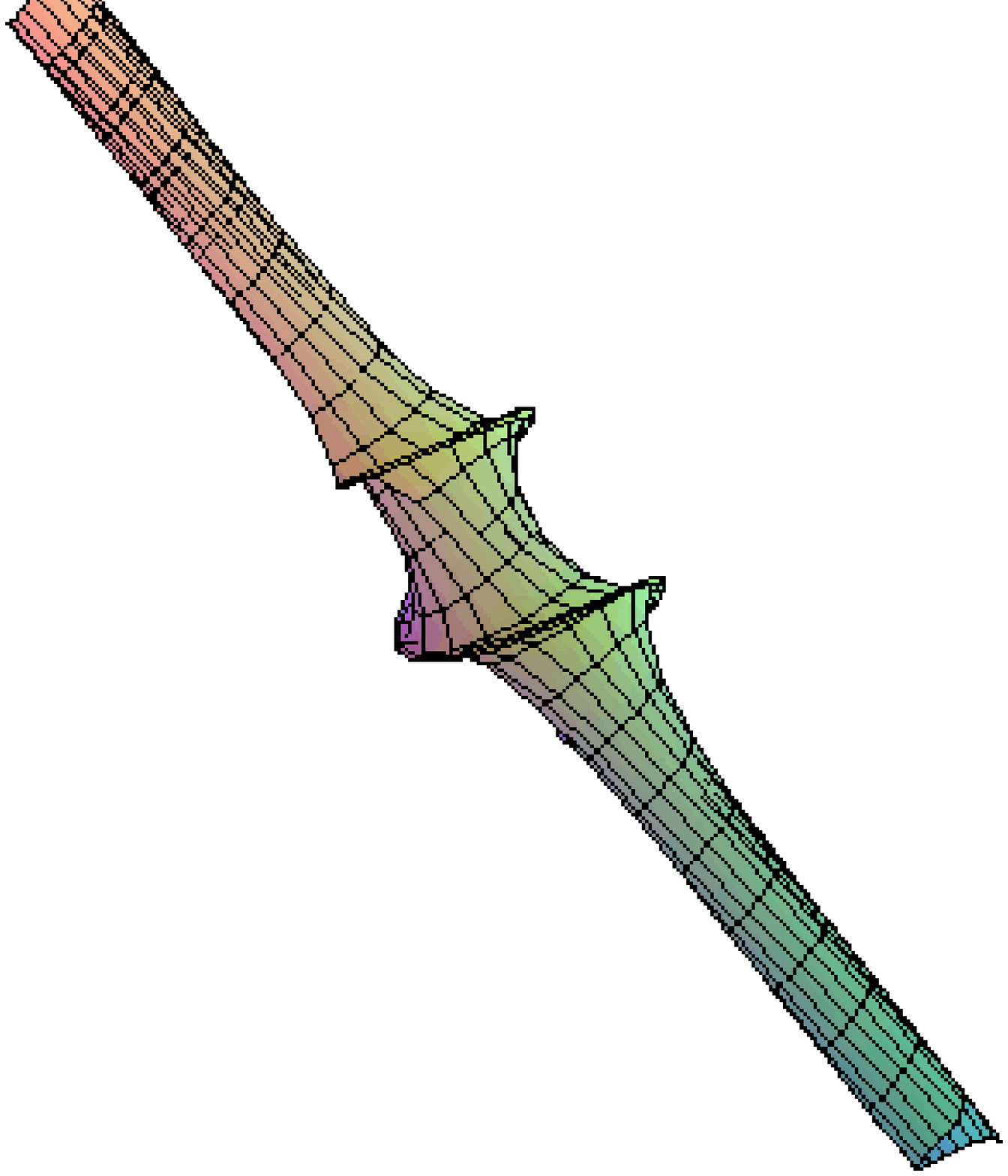}}
\caption{$(x,t)\in[-20,20]\times[-20,20]$}\label{fig7}
\end{figure}

The components of the position vector are
\begin{eqnarray}
&&y_{1}=-(25/13)\,{\textrm{sech}}\,{\xi}\,{{(e^{2\xi}-1)}/{(e^{2\xi}+1)}}
-{{E_{2}}}-8/({e^{2\xi}+1}),\\
&&y_{2}=\big[-4\,{\textrm{sech}}\,{\xi}-
(12/13)\,{{\big(e^{4\xi}+1\big)}/{(e^{2\xi}+1)^2}}+
(19/13)\,{\textrm{sech}}^2\,{\xi}\big]\cos{G}\nonumber\\
&&~~~~~~+(5/13)\,\sin{G}\,{\big(e^{2\xi}-1\big)}/{\big(e^{2\xi}+1\big)},
\end{eqnarray}
\begin{eqnarray}
&&y_{3}=\big[4\,{\textrm{sech}}\,{\xi}+
(12/13)\,{{\big(e^{4\xi}+1\big)}/{(e^{2\xi}+1)^2}}-
(19/13)\,{\textrm{sech}}^2\,{\xi}\big] \sin{G}\nonumber\\
&&~~~~~~+(5/13)\,\cos{G}\,{\big(e^{2\xi}-1\big)}/{\big(e^{2\xi}+1\big)},
\end{eqnarray}
where ${{E_{2}}}=13(20\,x+t)/250,$ $G=(47\,t-20\,x)/200$ and
$\xi=(4x+t)/8.$ As $\xi$ tends to $\pm\infty$, $y_{1}$ tends to
$\pm\infty$, $y_{2}$ approaches to $\cos{t}$ and $y_{3}$ to
$-\sin{t}.$

\vspace{0.3 cm}

\section{Conclusion}

In this work, we considered mKdV 2-surfaces by using two
deformations, spectral deformation and a combination of gauge and
spectral deformations of mKdV equation and its Lax pair. We found
the first and second fundamental forms, and the Gaussian and mean
curvatures of the corresponding surfaces. By solving the Lax
equation for a given solution of the mKdV equation and the
corresponding Lax pair, we also found position vectors of these
surfaces.

\noindent The surfaces arising from the spectral deformation are
Weingarten and Willmore-like surfaces. We also obtained some mKdV
surfaces from the variational principle for the Lagrange function,
that is a polynomial of the Gaussian and mean curvatures of the
surfaces corresponding to the spectral deformations of the Lax
pair of the mKdV equation. For some special values of parameters,
we plotted these three parameter family of surfaces in Examples
2-5.

\noindent In the case of the gauge-spectral parameter
deformations, we obtained a four parameter family of mKdV
surfaces. For some special values of the parameters in the
position vectors Eq. (\ref{y1SpecGau}) of these surfaces, we
plotted them in Examples 6-8.

\section{Acknowledments}
 I would like to thank Metin G{\" u}rses for his continuous help
 in this work. I would like also thank B. {\"O}zg{\"u}r
 Sar{\i}o{\u{g}}lu for his critical reading of the manuscript.
 This work is partially supported by the Scientific and
Technological Research Council of Turkey.

\end{document}